\documentclass{article}
\usepackage{spconf,amsmath,graphicx}

\usepackage{bm}
\usepackage{color}
\usepackage{pdfpages}
\usepackage{amssymb}
\usepackage{amsmath}
\usepackage{multirow}
\usepackage{hyperref}

\usepackage{algorithm}
\usepackage[noend]{algpseudocode}

\usepackage{pdfpages}

\title{Text2Poster: Laying out Stylized Texts on Retrieved Images}
\name{Chuhao Jin\textsuperscript{1,2}\qquad Hongteng Xu\textsuperscript{1,2}$^{\star}$\thanks{$\star$ Hongteng Xu and Ruihua Song are corresponding authors.}\qquad Ruihua Song\textsuperscript{1,2}$^{\star}$\qquad Zhiwu Lu\textsuperscript{1,2}\thanks{Our code is available at \url{https://github.com/chuhaojin/Text2Poster-ICASSP-22}.}}
\address{\textsuperscript{1}Gaoling School of Artificial Intelligence, Renmin University of China, Beijing 100872, China\\\textsuperscript{2}Beijing Key Laboratory of Big Data Management and Analysis Methods\\$\{$jinchuhao,~hongtengxu,~rsong$\}$@ruc.edu.cn}
\begin{document}
%
\maketitle
\begin{abstract}
Poster generation is a significant task for a wide range of applications, which is often time-consuming and requires lots of manual editing and artistic experience. 
In this paper, we propose a novel data-driven framework, called \textit{Text2Poster}, to automatically generate visually-effective posters from textual information. 
Imitating the process of manual poster editing, our framework leverages a large-scale pretrained visual-textual model to retrieve background images from given texts, lays out the texts on the images iteratively by cascaded auto-encoders, and finally, stylizes the texts by a matching-based method.
We learn the modules of the framework by weakly- and self-supervised learning strategies, mitigating the demand for labeled data. 
Both objective and subjective experiments demonstrate that our Text2Poster outperforms state-of-the-art methods, including academic research and commercial software, on the quality of generated posters.
\end{abstract}
\section{Introduction}
\label{sec:intro}

As a kind of media with both artistic and functional qualities, posters have been widely used in many commercial and non-commercial scenarios to advertise and spread information. 
For example, e-commercial platforms apply attractive posters to promote their items. 
The websites of social events like conferences are often decorated with fancy and informative posters. 
These high-quality posters are generated by embedding stylized texts into suitable background images, which depends on a lot of manual editing and non-quantitative artistic experience. 
However, such a time-consuming and subjective process cannot meet the huge and rapidly-increasing demands for well-designed posters in real-world applications, which reduces the efficiency of information diffusion and leads to sub-optimal promotion effects. 

\begin{figure*}[t]
    \centering
    \includegraphics[width=0.95\linewidth]{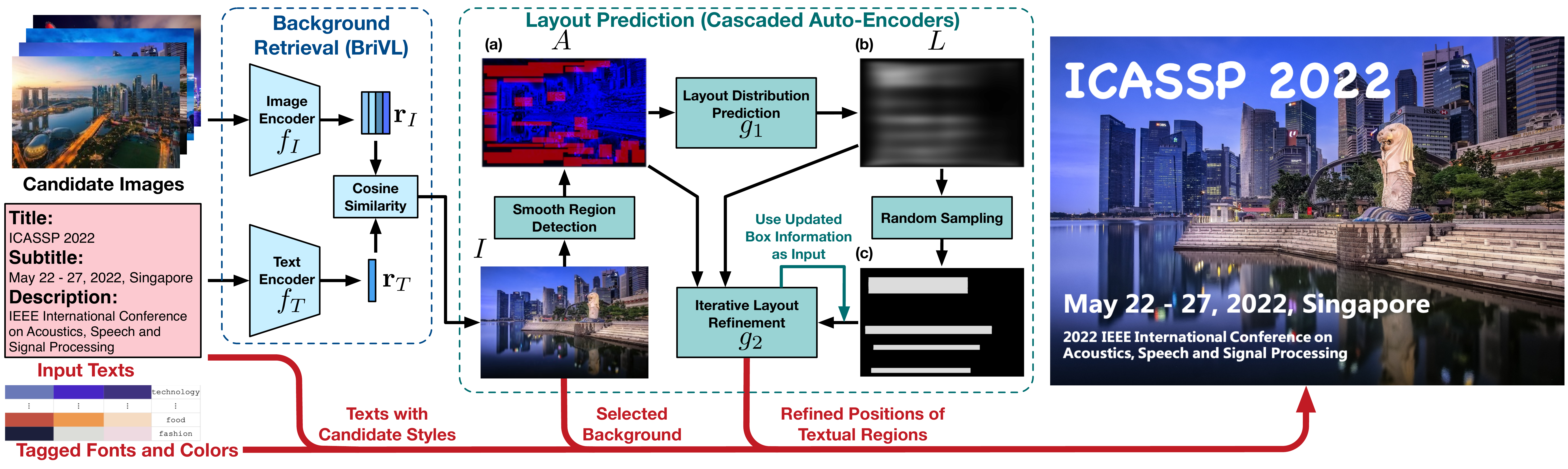}
    \vspace{-5pt}
    \caption{An illustration of our Text2Poster framework. \textbf{Note that the poster is generated by our method.}
    }
    \label{fig:scheme}
\end{figure*}

In this work, we propose a novel data-driven framework called \textit{Text2Poster}, which achieves a well-performed automatic poster generator. 
As illustrated in Fig.~\ref{fig:scheme}, the Text2Poster first leverages a large-scale pretrained visual-textual model to retrieve suitable background images from input texts. 
Then, this framework initializes the layout of the texts by sampling from the estimated layout distribution and refines the layout iteratively by cascaded auto-encoders.
Finally, it retrieves the color and font of the texts from a set of fonts and colors with semantic tags.
We learn the modules of the framework by applying weakly- and self-supervised learning strategies.
Experiments demonstrate that our Text2Poster framework can automatically generate high-quality posters, which outperforms its academic and industrial competitors greatly on both objective and subjective measurements.

\section{Related Work and Challenges}
As aforementioned, generating posters requires ($i$) determining background images from given textual information and ($ii$) optimizing the texts' layouts on the images. 
In the first step, most existing methods either do not leverage high-level features~\cite{huang2010review} or just consider the features from one side~\cite{peng2017image}, which often lead to undesired retrieval results.
Recently, the large-scale pretrained visual-textual models, such as Oscar~\cite{li2020oscar},  CLIP~\cite{radford2021learning}, and our BriVL~\cite{huo2021wenlan}, achieve encouraging performance on retrieving images from texts, which shows a potential solution to find background images for poster design. 
Note that although some methods make attempts to generate images from texts~\cite{mao2019bilinear,ramesh2021zero}, the quality of their generated images is unsuitable for poster generation.

For layout prediction, conventional rule-based methods~\cite{jahanian2013recommendation,yang2016automatic} explore texts' layouts from a limited number of predefined layout templates, whose flexibility is questionable. 
For the learning-based methods, the neural design network in~\cite{lee2020neural} requires side information like the regions of interests (ROIs) of background images to generate texts' layouts. 
The method in~\cite{li2020attribute} lays out each textual element sequentially while ignores the influences of the latter on the layout of the former. 
The LayoutGAN~\cite{li2018layoutgan}, LayoutGAN++~\cite{kikuchi2021constrained}, and SMARTTEXT~\cite{zhang2020smarttext} are developed for magazine or advertisement design. 
They do not fully consider the visual objects on background images when laying out texts and thus may lead to sub-optimal solutions to poster generation.

\section{Proposed Method}\label{sec:proposed}

Denote input texts as $\mathcal{T}=\{(T_i, \alpha_i)\}_{i=1}^{N_T}$, where $T_i$ represents the $i$-th textual information ($e.g.$, a sentence or a phrase) and $\alpha_i$ is the attribute of $T_i$ ($e.g.$, ``title'' or ``subtitle''). 
We generate a poster from $\mathcal{T}$ by the following three steps:
($i$) Retrieving a background image $I$ from a set of $N_I$ candidate images, denoted as $\mathcal{I}=\{I_j\}_{j=1}^{N_I}$. 
($ii$) Estimating the layout of each $T_i$, denoted as $\bm{P}=[\bm{p}_i]\in\mathbb{R}^{2\times N_T}$, where $\bm{p}_i\in\mathbb{R}^2$ represents the normalized top-left coordinate of $T_i$ on the background image $I$. 
($iii$) Determining each $T_i$'s font and color. 

\subsection{Image retrieval via a pretrained visual-textual model}
When retrieving background images for poster generation, we would like to explore the images ``weakly correlated'' with the texts.
For example, when retrieving images by the phrase ``The Wedding of Bob and Alice'', we prefer to find some images with love metaphors, $e.g.$, a picture showing a white church under the blue sky.
To achieve this aim, we leverage the BriVL we proposed in~\cite{huo2021wenlan}, one of the SOTA pretrained visual-textual models, to retrieve background images from texts. 
As shown in Fig.~\ref{fig:scheme}, BriVL consists of an image encoder $f_{T}$ and a text encoder $f_I$.
$f_T$ uses the encoder of RoBERTa-Large~\cite{liu2019roberta} as its textual backbone. 
$f_{I}$ uses a pretrained Faster R-CNN~\cite{ren2015faster} and an EfficientNet~\cite{tan2019efficientnet} as its visual backbone. 
Given the outputs of above backbone models, BriVL stacks four Transformer~\cite{vaswani2017attention} layers to derive 2048-dimensional visual and textual features. 
BriVL is trained on 30 million weakly-correlated image-text pairs from Internet and thus is suitable for our task, it applies the InfoNCE loss~\cite{oord2018representation} to align the features of texts to those of images.
Please refer the reader to~\cite{huo2021wenlan} for more its implementation details.

We collect 284,781 high-quality images from \url{unsplash.com} as our image retrieval library. 
Based on BriVL, we extract the latent codes of the input texts and the candidate images, $i.e.$, $\bm{r}_{\mathcal{T}}:=f_{T}(\cup_i T_i)$, and $\{\bm{r}_{I_j}:=f_{I}(I_j)\}_{j=1}^{N_I}$. 
Accordingly, we calculate the cosine similarity between $\bm{r}_{\mathcal{T}}$ and each $\bm{r}_{I_j}$ and retrieve the background image with the highest similarity, $i.e.$, $I=\arg\max_{I_j\in\mathcal{I}} \frac{\bm{r}_{\mathcal{T}}^{\top}\bm{r}_{I_j}}{\|\bm{r}_{\mathcal{T}}\|_2\|\bm{r}_{I_j}\|_2}$. 

\subsection{Layout prediction via cascaded auto-encoders}
Given the selected image $I$, we predict the layout of the input texts $\mathcal{T}$, $i.e.$, the $\bm{P}=[\bm{p}_i]$, by cascaded auto-encoders.

\textbf{Smooth Region Detection:} 
Inspired by the Faster R-CNN~\cite{ren2015faster}, we first generate $N_A$ overlapped regions with different sizes in $I$, denoted as $\{\mathcal{A}_i\}_{i=1}^{N_A}$. 
Applying the spectral residual approach~\cite{hou2007saliency}, we generate the saliency map of the background image $I$, denoted as $S$. 
For each region $\mathcal{A}_i$, we assign a value to it by calculating the averaged value of the saliency map in the region with a size-sensitive offset, $i.e.$, $v_i=\frac{1}{|\mathcal{A}_i|}(\lambda + \sum_{\bm{p}\in\mathcal{A}_i} S(\bm{p}))$, where $S(\bm{p})$ is the saliency at $\bm{p}$ and $|\mathcal{A}_i|$ is the number of pixels in $\mathcal{A}_i$. 
In principle, $\frac{1}{|\mathcal{A}_i|}\sum_{\bm{p}\in\mathcal{A}_i} S(\bm{p})$ is small when $\mathcal{A}_i$ is a smooth region, and the offset $\frac{\lambda}{|\mathcal{A}_i|}$ is small for large-sized $\mathcal{A}_i$. 
Therefore, we set a threshold $v_{\max}$ and select large regions with small values, $i.e.$, $\{\mathcal{A}_i | v_i<v_{\max}, \forall i=1,...,N_A\}$. 
For each image, we set $v_{\max}$ adaptively as $1.4\times \text{mean}\{v_i\}$ and apply the Non-Maximum Suppression (NMS) method~\cite{ren2015faster} to ensure the selected regions non-overlapped.  
The selected regions lead to a binary map indicating the smooth region of $I$, denoted as $A$.
The image (a) in Fig.~\ref{fig:scheme} shows the saliency map $S$ in blue and the smooth region map $A$ in red.

\textbf{Layout Distribution Prediction:} 
Given the map $A$, we leverage a generator $g_1$ to predict a layout distribution, denoted as $L$. 
For each pixel $\bm{p}$, $L(\bm{p})\in [0, 1]$ is proposed to indicate the probability that $\bm{p}$ belongs to a text box.
Here, $g_1$ owns an auto-encoding architecture, whose encoder $f_1$ is stacked CNNs and decoder $h_1$ is stacked Transposed-CNNs. 
Following the work in~\cite{li2020unicoder}, we construct the input of the decoder by concatenating the output of the encoder with a learnable position embedding map (denoted as $E$). 
Therefore, we have $L=g_1(A)=h_1(\text{Concat}(f_1(A), E))$.
The image (b) in Fig.~\ref{fig:scheme} illustrates the layout distribution $L$.

\textbf{Iterative Layout Refinement:} 
We treat $L$ as the prior of the target layout and initialize the layout $\bm{P}^{(0)}=[\bm{p}_i^{(0)}]$ by sampling each $\bm{p}_i^{(0)}$ from $L$. 
Taking unnormalized $\bm{p}_i^{(0)}$ as the top-left coordinate of the $i$-th text box, we initialize the box, whose size is determined by the length of the text $T_i$ and its attribute $\alpha_i$. 
The image (c) in Fig.~\ref{fig:scheme} illustrates the boxes. 

We leverage an auto-encoder, denoted as $g_2$, to refine the layout information in an auto-regressive manner, where
\begin{eqnarray}
    \bm{P}^{(k+1)} = 
    g_2(\text{Concat}(A, L), \bm{P}^{(k)}),~k=0,...,K-1
\end{eqnarray}
$K$ is the number of iterations. 
For $g_2$, its encoder is stacked CNNs, and its decoder is a 2-layer bidirectional LSTM. 

The auto-encoder for layout distribution prediction and that for layout refinement lead to a layout predictor with a cascaded auto-encoding architecture, which imitates the process of manual image editing. 
In practice, poster designers always avoid salient and informative regions when laying out texts. 
They often coarsely locate the texts and then adjust the positions iteratively. 
From this viewpoint, our layout predictor yields the same process to some degree.

\textbf{Learning Cascaded Auto-Encoders:} 
We train the two auto-decoders separately. 
In particular, we collect 154,013 poster images in 16 categories from \url{huaban.com}, a website encourages users to pin beautiful pictures. 
For each poster in the dataset, we first apply an OCR tool~\cite{du2020pp} to detect its texts $\widehat{\mathcal{T}}$ and the corresponding text boxes $\widehat{\bm{P}}=[\hat{\bm{p}}]$. 
The boxes lead to a binary layout map, denoted as $\widehat{L}$. 
Masking the image by the layout map and filling the masks by the image inpainting method in~\cite{telea2004image}, we obtain the background image $\widehat{I}$ of the poster. 
Applying the smooth region detector, we can obtain the smooth region map $\widehat{A}$ accordingly. 
As a result, we represent the dataset as $\mathcal{D}=\{(\widehat{\mathcal{T}}_n, \widehat{\bm{P}}_n, \widehat{L}_n, \widehat{A}_n)\}_{n=1}^{N}$. 
Accordingly, we train the two auto-encoders independently by
\begin{eqnarray}
\sideset{}{_{g_1}}\min  \frac{1}{N}\sideset{}{_{n=1}^{N}}\sum \|g_1(\widehat{A}_n)-\widehat{L}_n\|_2^2.\hspace{1.5cm}\\
\sideset{}{_{g_2}}\min \frac{1}{N}\sideset{}{_{n=1}^{N}}\sum
\|g_2(\text{Concat}(\widehat{A}_n,\widehat{L}_n),\bm{P}_{n}^{(0)}) - \widehat{\bm{P}}_{n}\|_F^2
\end{eqnarray}
where $|\widehat{\bm{P}}_n|$ indicates the number of boxes for the $n$-th poster. 
Here, we leverage a self-supervised learning strategy to train $g_2$, sampling the initial position $\bm{P}_{n}^{(0)}=[\bm{p}_{i,n}^{(0)}]$ by
$\bm{p}_{i,n}^{(0)} \sim \text{Uniform}(\hat{\bm{p}}_{i,n} - \Delta, \hat{\bm{p}}_{i,n} + \Delta)$, $\forall\hat{\bm{p}}_{i,n} \in \widehat{\bm{P}}_n$. 
The perturbation $\Delta=[0.1,0.1]^T$ controls the variance between the initial position and the target position. 
$\bm{P}_n^{(0)}$ is sampled based on the ground truth $\widehat{\bm{P}}_{n}$, achieving a self-supervised mechanism.

\textbf{Implementation Details:} 
Each convolution layer used in $g_1$ contains $16$ kernels with size $9\times 9$. 
The encoder of $g_1$ finally outputs a 64-dimensional feature vector. 
For the encoder of $g_2$, each of its convolution layers contains $64$ kernels with size $5\times 5$. 
For the 2-layer bidirectional LSTM (decoder) of $g_2$, the dimension of its hidden layer is set to be $200$. 
When training the two auto-encoders, we split the dataset $\mathcal{D}$ into 138,013 training posters and 16,000 validation posters. 
Each poster is resized to $300\times 400$. 
We leverage the Adam algorithm~\cite{kingma2015adam} to optimize the models with a learning rate of 0.05 and a batchsize of 512.
On four V100 GPUs, we train $g_1$ and $g_2$ for four and 48 hours, respectively. 

\begin{figure}[t]
    \centering
    \includegraphics[width=0.95\linewidth]{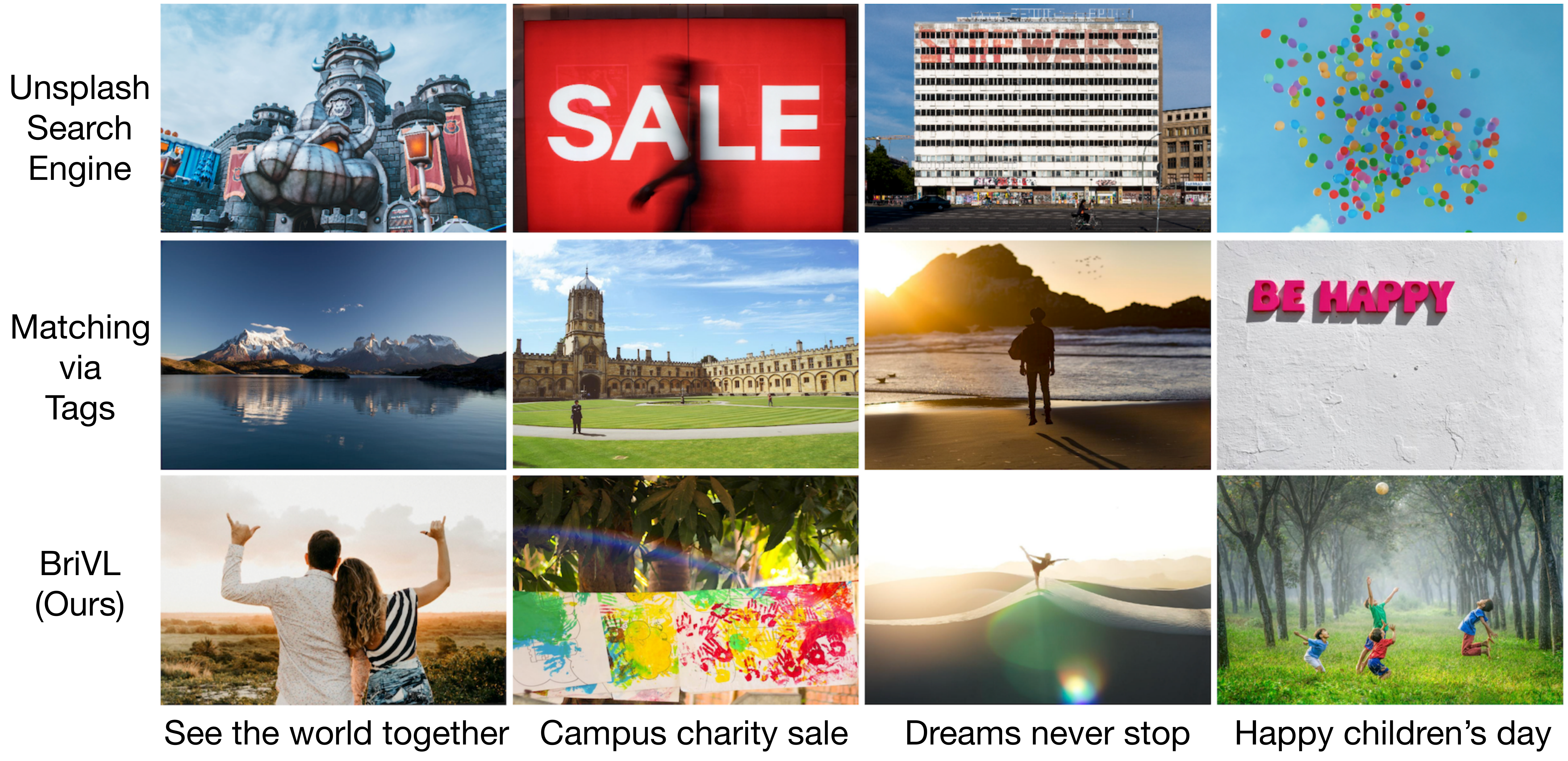}
    \vspace{-10pt}
    \caption{The images retrieved from different text queries.}
    \vspace{-10pt}
    \label{fig:eval}
\end{figure}

\subsection{Text stylizing}
Given the dataset $\mathcal{D}$, we construct a set of tuples, denoted as $\{(\widehat{\bm{r}}_{i,n}, c_{i,n}^I, c_{i,n}^T, z_{i,n})\}$, where $\widehat{\bm{r}}_{i,n}=f_T(\widehat{T}_{i,n})$ is the feature of the $i$-th text of $\widehat{\mathcal{T}}_n$, $c_{i,n}^I$ is the color of its background, $c_{i,n}^T$ is its color, and $z_{i,n}$ is its font. 
Given the tuples, we parse out 53 semantic tags according to the word frequency of the texts, and apply conventional clustering methods like K-means to find a set of typical textual styles corresponding to the tags, denoted as $\mathcal{F}=\{(\bm{r}_m, c_m^I, c_m^T, z_m)\}_{m=1}^{53}$. 
As a result, for each input text $T\in\mathcal{T}$, we extract its feature $\bm{r}=f_T(T)$ and obtain its background color as $c^I=I(\bm{p})$, where $\bm{p}$ is the predicted layout of $T$. 
Based on $(\bm{r},c^T)$, we find the matched clustering center from $\mathcal{F}$ under cosine similarity and determine the color and the font of $T$ accordingly.

\begin{figure*}[t]
    \centering
    \includegraphics[width=0.9\linewidth]{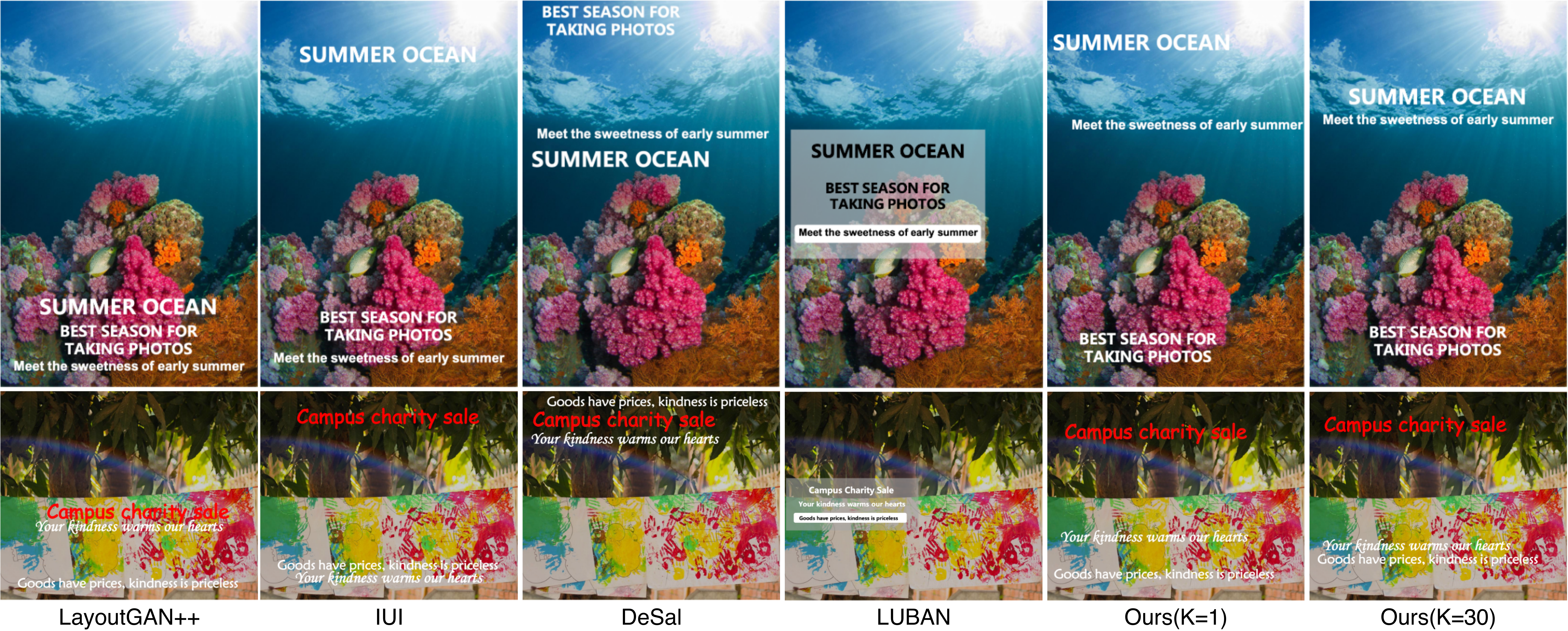}
    \vspace{-10pt}
    \caption{The posters generated by various layout prediction methods.}
    \label{fig:cmp}
\end{figure*}

\section{Experiments}\label{sec:experiments}
\textbf{Background Image Retrieval:}
Besides our BriVL-based image retrieval method, we consider ($a$) applying the search engine of \url{unsplash.com} and ($b$) matching the input texts with the tags of the images in our image retrieval library. 
Fig.~\ref{fig:eval} shows representative retrieval results obtained by different methods. 
The images retrieved by our method indeed contain the metaphors corresponding to the input texts. 
Given the text ``Campus charity sale'', the baselines tend to find images with explicit concepts like ``sale'' and ``campus'', while our method find an image with growing trees, rainbow, and colorful handprints, whose content reflects hidden but suitable semantics. 
Even for the challenging abstractive descriptions like ``See the world together'' and ``Dream never stop'', our method can still find suitable images.

In subjective evaluation, given 50 text queries, we retrieve top-5 images for each query by different methods. 
We invited three volunteers to score the quality of the retrieved image from 0 (very poor) to 4 (very well). 
The mean and the standard deviation of the scores are $2.17\pm 0.10$ for the Unsplash search engine, $1.64\pm 0.16$ for the tag-based matching method, and $\textbf{2.38}\pm 0.13$ for our BriVL-based method, which further demonstrates the superiority of our method.

\textbf{Layout Prediction:}
We evaluate our layout predictor quantitatively and qualitatively and compare it with the following baselines: 
($i$) the SOTA learning-based method \textit{LayoutGAN++}~\cite{kikuchi2021constrained}; 
($ii$) the SOTA rule-based methods \textit{IUI}~\cite{jahanian2013recommendation} and \textit{DeSal}~\cite{bylinskii2017learning};
($iii$) the commercial poster generator \textit{LUBAN} at \url{https://luban.aliyun.com}. 
To demonstrate the usefulness of our iterative layout refinement strategy, for our layout predictor, we set $K$ to be $1$, $5$, and $30$, respectively.

We construct a reference dataset by collecting 16,000 posters from \url{huaban.com} and prepare three background image sets: Unsplash2K, Unsplash10K, and PSD1.6K. 
Unsplash2K and Unsplash10K contain 2,000 and 10,000 background images from \url{unsplash.com}, respectively. 
PSD1.6K contains 1,637 background images extracted from the poster files in PSD format. 
For each image set, we lay out input texts by various methods on the background images and generate posters. 
We follow the work in~\cite{kikuchi2021constrained} and calculate the Fr\'{e}chet Inception Distance (FID) between the generated posters and the reference dataset. 
FID is widely used as the evaluation metric in GAN-based image generation tasks, it uses the Inception-v3 to measure the distribution distance between two datasets.
The results in Table~\ref{tab:obj_eval} show that our method outperforms the baselines consistently, and its performance is improved with the increase of $K$, which verifies the rationality of our iterative refinement strategy. 

Additionally, we manually select 50 text sets, each of which contains a title and several subtitles or descriptions. 
For each layout method, we first retrieve five background images from each text set by the BriVL model and generate 250 posters accordingly. 
We invite 13 volunteers to score these generated posters from 0 (very poor) to 4 (very well) on the aesthetics of the layouts. 
For each method, the mean and the standard deviation of the scores are shown in Table~\ref{tab:obj_eval}. 
Our method also achieves the best result in the subjective experiment.
Fig.~\ref{fig:cmp} shows some generated posters, which further verify the effectiveness of our Text2Poster method.

\begin{table}[t]
\centering
\begin{small}
\begin{tabular}{@{\hspace{1pt}}c@{\hspace{2pt}}|@{\hspace{2pt}}c@{\hspace{3pt}}c@{\hspace{3pt}}c|c@{\hspace{1pt}}}
\hline
    \multirow{2}{*}{Method} & \multicolumn{3}{c|}{FID} & Aesthetics\\
    & PSD1.6K & Unsplash2K & Unsplash10K & Score\\ \hline
    LayoutGAN++ & 
    55.46  & 
    75.63  & 
    58.11  &   
    1.63$_{\pm\text{0.22}}$ \\
    IUI & 
    44.39  & 
    68.85  & 
    53.24  &
    2.25$_{\pm\text{0.24}}$ \\
    DeSal  & 
    43.47  & 
    70.30  & 
    55.25  &
    2.28$_{\pm\text{0.32}}$ \\
    Ours ($K=1$) & 
    41.83  & 
    71.64  & 
    55.20  &
    2.14$_{\pm\text{0.25}}$ \\
    Ours ($K=5$)  & 
    39.28  & 
    68.77  & 
    52.78  &             
    2.36$_{\pm\text{0.21}}$ \\
    Ours ($K=30$) & 
    \textbf{39.09}  & 
    \textbf{67.94}  & 
    \textbf{52.75}  &  
    \textbf{2.39}$_{\pm\textbf{0.22}}$ \\
    \hline
\end{tabular}
\end{small}
\caption{Objective and subjective evaluations of various layout prediction methods. 
\textit{LUBAN only provides charged services and thus is unavailable for large-scale numerical test.}}
\label{tab:obj_eval}
\end{table}

\section{Conclusions}\label{sec:conclusions}

In this paper, we propose a novel framework to generate posters from input texts in an automatic way, which achieves state-of-the-art performance.
We take advantage of a pretrained visual-textual model for image retrieval and cascaded autoencoders for layout prediction. 
The modules of the framework is trained by cutting-edge weakly- and self-supervised learning strategies. 
In the future, we plan to improve the framework by fine-tuning the pretrained BriVL model and applying an end-to-end learning strategy.

\section{Acknowledgements}\label{sec:acknowledgements}
This work was supported by Beijing Outstanding Young Scientist Program NO. BJJWZYJH012019100020098, Large-Scale Pre-Training Program 468 of Beijing Academy of Artificial Intelligence, Beijing Key Laboratory of Big Data Management and Analysis Methods, and Intelligent Social Governance Platform, Major Innovation \& Planning Interdisciplinary Platform for the ``Double-First Class'' Initiative of RUC. We also wish to acknowledge the support provided by Public Policy and Decision-making Research Lab of RUC.

\newpage
\bibliographystyle{IEEEbib}
\bibliography{refs_compressed.bib}

\begin{thebibliography}{10}

\bibitem{huang2010review}
Wei Huang, Yan Gao, and Kap~Luk Chan,
\newblock ``A review of region-based image retrieval,''
\newblock {\em Journal of Signal Processing Systems}, vol. 59, no. 2, pp.
  143--161, 2010.

\bibitem{peng2017image}
Tian-qiang Peng and Fang Li,
\newblock ``Image retrieval based on deep convolutional neural networks and
  binary hashing learning,''
\newblock in {\em ICASSP}, 2017, pp. 1742--1746.

\bibitem{li2020oscar}
Xiujun Li, Xi~Yin, Chunyuan Li, Pengchuan Zhang, Xiaowei Hu, Lei Zhang, Lijuan
  Wang, Houdong Hu, et~al.,
\newblock ``Oscar: Object-semantics aligned pre-training for vision-language
  tasks,''
\newblock in {\em ECCV}, 2020, pp. 121--137.

\bibitem{radford2021learning}
Alec Radford, Jong~Wook Kim, Chris Hallacy, Aditya Ramesh, Gabriel Goh,
  Sandhini Agarwal, Girish Sastry, Amanda Askell, Pamela Mishkin, et~al.,
\newblock ``Learning transferable visual models from natural language
  supervision,''
\newblock {\em arXiv preprint arXiv:2103.00020}, 2021.

\bibitem{huo2021wenlan}
Yuqi Huo, Manli Zhang, Guangzhen Liu, Haoyu Lu, Yizhao Gao, Guoxing Yang,
  Jingyuan Wen, Heng Zhang, Baogui Xu, et~al.,
\newblock ``Wenlan: Bridging vision and language by large-scale multi-modal
  pre-training,''
\newblock {\em arXiv preprint arXiv:2103.06561}, 2021.

\bibitem{mao2019bilinear}
Xiaofeng Mao, Yuefeng Chen, Yuhong Li, Tao Xiong, et~al.,
\newblock ``Bilinear representation for language-based image editing using
  conditional generative adversarial networks,''
\newblock in {\em ICASSP}, 2019, pp. 2047--2051.

\bibitem{ramesh2021zero}
Aditya Ramesh, Mikhail Pavlov, Gabriel Goh, Scott Gray, Chelsea Voss, Alec
  Radford, Mark Chen, and Ilya Sutskever,
\newblock ``Zero-shot text-to-image generation,''
\newblock {\em arXiv preprint arXiv:2102.12092}, 2021.

\bibitem{jahanian2013recommendation}
Ali Jahanian, Jerry Liu, Qian Lin, Daniel Tretter, Eamonn O'Brien-Strain,
  Seungyon~Claire Lee, Nic Lyons, and Jan Allebach,
\newblock ``Recommendation system for automatic design of magazine covers,''
\newblock in {\em IUI}, 2013, pp. 95--106.

\bibitem{yang2016automatic}
Xuyong Yang, Tao Mei, Ying-Qing Xu, Yong Rui, and Shipeng Li,
\newblock ``Automatic generation of visual-textual presentation layout,''
\newblock {\em TOMM}, vol. 12, no. 2, pp. 1--22, 2016.

\bibitem{lee2020neural}
Hsin-Ying Lee, Lu~Jiang, Irfan Essa, Phuong~B Le, Haifeng Gong, Ming-Hsuan
  Yang, and Weilong Yang,
\newblock ``Neural design network: Graphic layout generation with
  constraints,''
\newblock in {\em ECCV}, 2020, pp. 491--506.

\bibitem{li2020attribute}
Jianan Li, Jimei Yang, Jianming Zhang, Chang Liu, Christina Wang, and Tingfa
  Xu,
\newblock ``Attribute-conditioned layout gan for automatic graphic design,''
\newblock {\em TVCG}, 2020.

\bibitem{li2018layoutgan}
Jianan Li, Jimei Yang, Aaron Hertzmann, Jianming Zhang, and Tingfa Xu,
\newblock ``Layoutgan: Generating graphic layouts with wireframe
  discriminators,''
\newblock in {\em ICLR}, 2018.

\bibitem{kikuchi2021constrained}
Kotaro Kikuchi, Edgar Simo-Serra, Mayu Otani, and Kota Yamaguchi,
\newblock ``Constrained graphic layout generation via latent optimization,''
\newblock {\em arXiv preprint arXiv:2108.00871}, 2021.

\bibitem{zhang2020smarttext}
Peiying Zhang, Chenhui Li, and Changbo Wang,
\newblock ``Smarttext: Learning to generate harmonious textual layout over
  natural image,''
\newblock in {\em ICME}, 2020, pp. 1--6.

\bibitem{liu2019roberta}
Yinhan Liu, Myle Ott, Naman Goyal, Jingfei Du, Mandar Joshi, Danqi Chen, Omer
  Levy, Mike Lewis, Luke Zettlemoyer, and Veselin Stoyanov,
\newblock ``Roberta: A robustly optimized bert pretraining approach,''
\newblock {\em arXiv preprint arXiv:1907.11692}, 2019.

\bibitem{ren2015faster}
Shaoqing Ren, Kaiming He, Ross Girshick, and Jian Sun,
\newblock ``Faster r-cnn: Towards real-time object detection with region
  proposal networks,''
\newblock in {\em NeurIPS}, 2015, pp. 91--99.

\bibitem{tan2019efficientnet}
Mingxing Tan and Quoc Le,
\newblock ``Efficientnet: Rethinking model scaling for convolutional neural
  networks,''
\newblock in {\em ICML}, 2019, pp. 6105--6114.

\bibitem{vaswani2017attention}
Ashish Vaswani, Noam Shazeer, Niki Parmar, Jakob Uszkoreit, Llion Jones,
  Aidan~N Gomez, {\L}ukasz Kaiser, and Illia Polosukhin,
\newblock ``Attention is all you need,''
\newblock in {\em NeurIPS}, 2017, pp. 5998--6008.

\bibitem{oord2018representation}
Aaron van~den Oord, Yazhe Li, and Oriol Vinyals,
\newblock ``Representation learning with contrastive predictive coding,''
\newblock {\em arXiv preprint arXiv:1807.03748}, 2018.

\bibitem{hou2007saliency}
Xiaodi Hou and Liqing Zhang,
\newblock ``Saliency detection: A spectral residual approach,''
\newblock in {\em CVPR}, 2007, pp. 1--8.

\bibitem{li2020unicoder}
Gen Li, Nan Duan, Yuejian Fang, Ming Gong, and Daxin Jiang,
\newblock ``Unicoder-vl: A universal encoder for vision and language by
  cross-modal pre-training,''
\newblock in {\em AAAI}, 2020, vol.~34, pp. 11336--11344.

\bibitem{du2020pp}
Yuning Du, Chenxia Li, Ruoyu Guo, Xiaoting Yin, Weiwei Liu, Jun Zhou, Yifan
  Bai, Zilin Yu, Yehua Yang, et~al.,
\newblock ``Pp-ocr: A practical ultra lightweight ocr system,''
\newblock {\em arXiv preprint arXiv:2009.09941}, 2020.

\bibitem{telea2004image}
Alexandru Telea,
\newblock ``An image inpainting technique based on the fast marching method,''
\newblock {\em Journal of Graphics Tools}, vol. 9, no. 1, pp. 23--34, 2004.

\bibitem{kingma2015adam}
Diederik~P Kingma and Jimmy Ba,
\newblock ``Adam: A method for stochastic optimization,''
\newblock in {\em ICLR}, 2015.

\bibitem{bylinskii2017learning}
Zoya Bylinskii, Nam~Wook Kim, Peter O'Donovan, Sami Alsheikh, Spandan Madan,
  Hanspeter Pfister, Fredo Durand, Bryan Russell, and Aaron Hertzmann,
\newblock ``Learning visual importance for graphic designs and data
  visualizations,''
\newblock in {\em ACM UIST}, 2017, pp. 57--69.

\end{thebibliography}

\end{document}